\documentclass[aps,prl,amsmath,twocolumn]{revtex4-1}

\usepackage{amsmath}
\usepackage{bm}
\usepackage[dvips]{color,graphicx}
\usepackage{epsfig}
\usepackage{dcolumn}
\usepackage{latexsym}
\usepackage{dcolumn}

\def\sintwms{\sin^2\hat\theta_W}
\def\cs{^{133}{\rm Cs}}
\def\ra{^{213}{\rm Ra}}
\def\swms{\hat{s}^2}

\def\vems{\hat{v}_e}

\def\gz{{\gamma Z}}

\begin{document}

\title{$\gz$ box corrections to weak charges of heavy nuclei
	in atomic parity violation}

\author{P.~G.~Blunden,$^1$
	W.~Melnitchouk$^2$ and
	A.~W.~Thomas$^3$}
\affiliation{
$^1$\mbox{Department of Physics and Astronomy, University of Manitoba}, 
	Winnipeg, MB, Canada R3T 2N2	\\
$^2$\mbox{Jefferson Lab, 12000 Jefferson Avenue, Newport News,
	Virginia 23606, USA}		\\
$^3$\mbox{CSSM and CoEPP, School of Chemistry and Physics, University of 
	Adelaide}, Adelaide SA 5005, Australia}

\begin{abstract}
We present a new dispersive formulation of the $\gz$ box radiative
corrections to weak charges of bound protons and neutrons in atomic
parity violation (APV) measurements on heavy nuclei such as $\cs$ and
$\ra$.  We evaluate for the first time a small but important additional
correction arising from Pauli blocking of nucleons in a heavy nucleus.
Overall, we find a significant shift in the $\gz$ correction to the
weak charge of $\cs$, approximately 4 times larger than the current
uncertainty on the value of $\sintwms$, but with a reduced error
compared to earlier estimates.
\end{abstract}


\maketitle

In the search for physics beyond the Standard Model one of the most
important indirect methods involves a high precision test of the
evolution of the Weinberg angle with scale.  In particular, the
comparison between the value of $\sintwms$ measured at the $Z$ pole
at LEP and the value extracted from parity violation in atomic systems
has provided a very strong confirmation of the radiative corrections
calculated within the Standard Model.
The atomic system with the most accurate current measurement is the
parity-violating $S$--$S$ transition in neutral Cs \cite{Wood97,Porsev09}.
The predicted enhancement factor of the APV effect in the $S$--$D$
transition in Ra$^+$ is about 50 times larger~\cite{Dzuba01}.
It is critical that the calculations of the radiative corrections for
these systems incorporate the latest theoretical developments and match
the precision of the experimental data.

Driven by the demand for accurate radiative corrections for high 
energy parity-violating electron scattering, notably the
Q$_{\rm weak}$ experiment at Jefferson Lab \cite{Qweak}, there
have recently been new evaluations of the $\gz$ box diagram
\cite{GH09,SBMT10,RC11,GHRM11,BMT11}.
In particular, Blunden {\it et al.}~\cite{BMT11} developed a
formulation of this correction in terms of dispersion relations
and moments of the inclusive $\gz$ interference  structure
functions.  This approach provides a systematic method for improving
the accuracy of the calculation.

In this Letter we use the dispersive relations methods to compute a
new value for the radiative correction associated with the $\gz$
box diagram for both protons and neutrons in $\cs$ and $\ra$.
Compared with previous estimates that were computed some 3 decades
ago~\cite{MS}, the new corrections give $\gz$ contributions that are
15\% smaller for free nucleons, and 20\% smaller for nucleons bound
in a heavy nucleus (due to Pauli blocking).
This shifts the theoretical value of $Q_W(\rm{Cs})$ from $-73.14(6)$
to $-73.26(4)$, which agrees well with the current experimental value
of $-73.20(35)$.

Including radiative corrections, the weak charges of the proton and
neutron can be written as~\cite{MS,Erler03}
\begin{eqnarray}
Q_W^p &=& (\rho + \Delta_e)
	  (1 - 4 \kappa(0)\swms + \Delta_e') \nonumber\\
      && +\, \Box_{WW}^p + \Box_{ZZ}^p
      + \Box_\gz^p,\label{eq:Qwp}\\
Q_W^n &=& - (\rho + \Delta_e)
      + \Box_{WW}^n + \Box_{ZZ}^n
      + \Box_\gz^n,\label{eq:Qwn}
\end{eqnarray}
with $\swms\equiv\sintwms(M_Z^2)=0.23116(13)$ in the
$\overline{\rm MS}$ scheme. $\Delta_e$ and $\Delta_e'$ are $Zee$ and
$\gamma ee$ vertex corrections, given in Refs.~\cite{MS,Erler03},
and expressions for the universal parameters $\rho$ and $\kappa(Q^2)$
are given in Ref.~\cite{Czar00}.

The proton weak charge $Q_W^p$ is sensitive to the running of the
Weinberg angle, as embodied in $\kappa(Q^2)$, whereas the neutron
weak charge $Q_W^n$ is primarily sensitive to $\rho$.  At the one-loop
level, $\rho$ has a quadratic dependence on the top quark mass.
This dependence is modified by significant higher-order QCD and
electroweak corrections~\cite{Faisst}, some of which have been 
evaluated to four loops.

The correction $\kappa(Q^2)$ includes boson self-energy contributions
from $\gz$ mixing.  In particular, it has a hadronic uncertainty
from the quark contributions to fermion loops, denoted by
$\Delta\kappa_{\rm had}^{(5)}$ for five quark flavors, which
is strongly correlated with the analogous contribution
$\Delta\alpha_{\rm had}^{(5)}$ to the running of $\alpha(Q^2)$.
A reduction of $\Delta\alpha_{\rm had}^{(5)}$ to 0.02772 in the
most recent analysis implies a corresponding reduction in
$\Delta\kappa_{\rm had}^{(5)}$ compared with Ref.~\cite{Erler03},
leading to $\Delta\kappa_{\rm had}^{(5)}\swms = 7.87(8) \times 10^{-3}$.
Using the most recent Standard Model parameters \cite{PDG12},
we obtain $\rho = 1.0006(2)$ and $\kappa(0)\swms = 0.23807(15)$,
in essential agreement with the values quoted in Ref.~\cite{PDG12}.

The remaining terms in Eqs.~(\ref{eq:Qwp}) and (\ref{eq:Qwn}) arise
from the (numerically dominant) $WW$, $ZZ$, and $\gz$ boxes.
We take the expressions for $WW$ and $ZZ$ boxes from Erler {\em et al.}
\cite{Erler03}, including the leading order QCD correction to the
original expressions of Ref.~\cite{MS}.
The box diagrams with two heavy bosons are dominated by high momentum
scales \cite{Marciano93}.  By contrast, the $\gz$ box diagram contains
both high and low momentum scales, and is therefore sensitive to
hadronic corrections.
Following convention we write the $\gz$ contribution for protons and
neutrons in terms of a parameter $B^N$,
\begin{equation}
\Box_{\gz}^N={3 \alpha\over 2\pi} \vems
\,{\cal Q}^N B^N\, ,
\end{equation}
where $\vems\equiv 1-4\swms$ and ${\cal Q}^N = 5/3\,(4/3)$
for the proton (neutron).
A free-quark model gives $B^p = \ln(M_Z^2/m^2) + 3/2$,
where $m$ is an undetermined hadronic mass scale (such as a constituent
quark mass); however, this does not adequately describe either the
long-range or short-range behavior of the $\gz$ box.
Marciano and Sirlin (MS) \cite{MS} give a more refined estimate by
separately modeling the low and high energy contributions, but an
equivalent logarithmic dependence on the hadronic mass parameter
$m$ remains.

Using Eqs.~(\ref{eq:Qwp}) and (\ref{eq:Qwn}), we find numerically
\begin{eqnarray}
Q_W^p &=& ~~0.0664(6) + 0.00044\, B^p,\\
Q_W^n &=& -0.9922(2) + 0.00035\, B^n.
\label{eq:Qwnum}
\end{eqnarray}
The error in $Q_W^p$ (excluding $\gz$ boxes) arises from
$\kappa(0) \swms$ ($\pm 0.0006$) and the $WW$ boxes ($\pm 0.0001$),
while the error in $Q_W^n$ arises from $\rho$ ($\pm 0.0002$).
The results in the column labeled MS in Table~\ref{tab:summary}
use the most recent estimates of $B^p$ and $B^n$ from
Refs.~\cite{Marciano93,Erler03}.
For $\cs$, we have $Q_W({\rm Cs}) = 55 Q_W^p + 78 Q_W^n$.
Adding the independent errors in quadrature (the errors in $B^p$ and
$B^n$ are not independent), we find $-73.14(6)$.  The $\gz$ boxes
contribute $\pm 0.052$ to this total, while all other errors combined
contribute $\pm 0.037$.
(See also the recent review of APV, including $\cs$, in Ref.~\cite{Dzuba12}.)

\begin{table}[t]
\vspace{-6pt}
\begin{center} 
\caption{
  Weak charges of the proton, neutron, $\cs$, and $\ra$, comparing the
  previous MS estimates~\cite{MS,Marciano93} with our new results for
  free and bound nucleons.
\label{tab:summary}
}
\begin{ruledtabular}
\begin{tabular}{lccc}
	&\text{MS} & \text{free nucleons} & \text{bound nucleons}\\ \hline
$B^p$	& 11.8(1.0) & 9.95(40) & 9.36(40)\\
$B^n$	& 11.5(1.0) & 9.82(40) & 9.32(40)\\ \hline
$Q_W^p$	& ~0.0716(8) & ~0.0708(6) & ~0.0705(6) \\
$Q_W^n$	& $-0.9882(4)$ & $-0.9888(2)$ & $-0.9890(2)$\\
$Q_W({\rm Cs})$
	& $-73.14(6)$ & $-73.23(4)$ & $-73.26(4)$\\
$Q_W({\rm Ra})$
	& $-117.22(10)$ & $-117.37(7)$~~ & $-117.42(7)$~~\\
\end{tabular}
\end{ruledtabular}
\end{center}
\vspace{-12pt}
\end{table}

To proceed beyond the work of MS~\cite{MS} we use the dispersion
methods developed in Refs.~\cite{GH09,SBMT10,RC11,GHRM11,BMT11},
writing the imaginary part of $\Box_\gz(E)$ in terms of structure
functions $F_{i,\gz}^N$ ($i=1,2,3$) that can be obtained from
inclusive lepton-nucleon scattering.
A dispersion integral over energy then gives the real part
of $\Box_\gz(E)$, which contributes to the weak charge.
The energy dependence of $\Box_\gz(E)$ was evaluated in
Refs.~\cite{SBMT10,BMT11}.
At $E=0$, relevant for APV, only the axial-vector $ZN$ coupling
involving the $F_{3,\gz}^N$ structure function gives a nonzero
contribution.  Performing the dispersion energy integral
analytically, the real part of $\Box_\gz$ can be written
\begin{eqnarray}
\hspace{-1em}
\Box_\gz^N
&=& {2\over\pi}
    \int_{0}^\infty dQ^2
    {\alpha(Q^2)\, v_e(Q^2)\over Q^2(1+Q^2/M_Z^2)}	 \nonumber\\
& & \hspace{-.5em}
    \times
    \int_0^1\!\! dx\ F_{3,\gz}^N(x,Q^2)\,
    \frac{1+2\gamma}{(1+\gamma)^2},
\label{eq:F3dis}
\end{eqnarray}
where $\gamma = \sqrt{1+4 M^2 x^2/Q^2}$ and $x = Q^2/(W^2-M^2+Q^2)$,
with $W$ the invariant mass of the intermediate hadronic state.
Following Ref.~\cite{BMT11}, we include the running with $Q^2$ of
$\alpha(Q^2)$ and $v_e(Q^2) \equiv 1-4 \kappa(Q^2)\, \hat{s}^2$
in Eq.~(\ref{eq:F3dis}) due to boson self-energy contributions.
Both quantities vary significantly over the relevant $Q^2$ range.

The contributions to $\Box_\gz$ can be split into three kinematic
regions:
(i) elastic, with $W^2=M^2$;
(ii) resonances, with $(M+m_\pi)^2\le W^2 \lesssim 4$~GeV$^2$; and
(iii) deep-inelastic scattering (DIS), with $W^2>4$~GeV$^2$.
Contributions from region (i) depend on the nucleon's elastic
magnetic $G_M^N$ and axial-vector $G_A^{Z,N}$ form factors,
\begin{equation}
F_{3,\gz}^{N(\rm el)}(x,Q^2)
= - G_M^N(Q^2) G_A^{Z,N}(Q^2)\, x\, \delta(1-x).
\end{equation}
We set
	$G_M^N(Q^2) = \mu^N F_V(Q^2)$, and
	$G_A^{Z,N}(Q^2) = -g_A^N F_A(Q^2)$,
with $\mu^N$ the nucleon magnetic moment, and
	$g_A^p = -g_A^n = 1.267$.
A dipole $Q^2$ dependence,
	$F_{V,A}(Q^2) = 1/(1+Q^2/\Lambda_{V,A}^2)^2$,
suffices for both $V$ and $A$ form factors, with $\Lambda_V=0.84$~GeV,
and $\Lambda_A=1.0$~GeV.  More sophisticated form factors give
essentially identical numerical results.

For the resonance contributions we use the parametrizations of the
transition form factors from Lalakulich {\it et al.}~\cite{LPP},
but with modified isospin factors appropriate to $\gz$.
These form factors have been fit to pion production data in $\nu$
and $\bar\nu$ scattering, and include the lowest four spin-1/2 and
3/2 states.

For the DIS region, we divide the $Q^2$ integral of Eq.~(\ref{eq:F3dis})
into a low-$Q^2$ part $Q^2 < Q_0^2$, where the structure function
$F_{3,\gz}^N$ is relatively unknown, and a high-$Q^2$ part
($Q^2 > Q_0^2$), where at leading order the structure function can be
expressed in terms of valence quark distributions~\cite{PDG12}.
At high $Q^2$ the $\gz$ contribution can be expanded in powers of
$x^2/Q^2$, yielding a series whose coefficients are structure
function moments of increasing rank,
\begin{eqnarray}  
\hspace{-1em}
\Box_\gz^{\rm (DIS)}
&=& {3\over 2\pi}
    \int_{Q_0^2}^\infty dQ^2
    {\alpha(Q^2)\, v_e(Q^2)\over Q^2(1+Q^2/M_Z^2)}  \nonumber\\
& & \times
    \biggl[ M_3^{(1)}(Q^2)
	  - {2M^2 \over 3Q^2} M_3^{(3)}(Q^2) + \ldots
    \biggr],
\label{eq:moments}
\end{eqnarray}
where the $n$-th moment of the $F_3^\gz$ structure function is
$M_3^{(n)}(Q^2) = \int_0^1 dx\,x^{n-1} F_3^\gz(x,Q^2)$.
Numerically, the $n=1$ moment dominates, with the $n\ge 3$ contributions
to the integral of Eq.~(\ref{eq:moments}) less than 0.1\%.

The lowest moment is in fact the $\gz$ analog of the GLS sum rule
\cite{GLS} for $\nu N$ DIS, which at leading order counts the number
of valence quarks in the nucleon.  The corresponding quantity for
$\gz$ is ${\cal Q}^p = \sum_q 2\, e_q\, g_A^q = 5/3$ for the proton
and ${\cal Q}^n = 4/3$ for the neutron.  Including the next-to-leading
order strong interaction correction in the $\overline{\rm MS}$ scheme,
the $n=1$ contribution is
\begin{equation}
M_3^{N(1)}(Q^2)
= {\cal Q}^N \left( 1 - {\alpha_s(Q^2)\over \pi} \right).
\end{equation}
Combined with Eq.~(\ref{eq:moments}), this is identical to the high
energy result of MS \cite{MS}, but with $Q_0$ replacing the
arbitrary hadronic mass parameter $m$.  In our case the scale $Q_0$
corresponds to the momentum above which a partonic representation of
the nonresonant structure functions is valid, and above which the
$Q^2$ evolution of parton distribution functions (PDFs) {\it via}
the $Q^2$ evolution equations is applicable.  We vary $Q_0^2$ between
1 and 2~GeV$^2$, which coincides with the typical lower limit of
recent sets of PDFs.

The contribution for $Q^2 < Q_0^2$ can in principle be obtained
from data.  There is limited information on $F_{3,W}$ from neutrino
scattering, but little or no existing data on $F_{3,\gz}$ at low $Q^2$.
As in Ref.~\cite{BMT11}, we use two different models to smoothly
interpolate in $Q^2$ between $Q_0^2$ and 0: one vanishes in the
$Q^2 \to 0$ limit, and the other approaches a constant.
The differences between the models are of the order 5--15\%.

The relation between proton and neutron contributions from the
different kinematic regions is
\begin{eqnarray}
\Box_\gz^{n{\rm (el)}}&=&-{\mu^n\over \mu^p} \Box_\gz^{p{\rm (el)}},\qquad
\Box_\gz^{n{\rm (res)}}\approx \Box_\gz^{p{\rm (res)}},\nonumber\\
\Box_\gz^{n{\rm (DIS)}}&=&{4\over 5} \Box_\gz^{p{\rm (DIS)}}.
\label{eq:pton}
\end{eqnarray}
The near equality of the resonance contributions (within 3\%) is due
to the dominance of isovector resonances, which contribute equally for
protons and neutrons.

\begin{table}[b]
\vspace{-6pt}
\begin{center} 
\caption{
  Contributions to $B^p$ from different kinematic regions for
  two different values of the matching scale $Q_0^2$.
  The range of values for the DIS ($Q^2<Q_0^2$) contribution is for the 
  two models described in Ref.~\cite{BMT11}.
  \label{tab:boxa}
}
\begin{ruledtabular}
\begin{tabular}{lcc}	
$Q_0^2$		& 1~GeV$^2$    	& 2~GeV$^2$	\\ \hline
elastic         & 1.47       	& 1.47 	\\
resonance       & 0.59        	& 0.59 	\\
DIS					\\
~~$Q^2>Q_0^2$	& 7.50 		& 7.05	\\
~~$Q^2<Q_0^2$	& 0.42--0.48	& 0.78--0.82 \\ \hline
Total		& 9.98--10.04	& 9.89--9.93 \\
\end{tabular}
\end{ruledtabular}
\end{center}
\vspace{-12pt}
\end{table}

The full results are summarized in Table~\ref{tab:boxa}.
The largest contributions come from the DIS region.
Fortunately, the results show only a mild sensitivity to the parameter
$Q_0$ that separates the model dependent low-$Q^2$ extrapolation from
the high-$Q^2$ partonic region.  The main part of the uncertainty
arises from the model dependence of this extrapolation to low $Q^2$.
We therefore assign $B^p=9.95(40)$, equal to the average of the four 
values in Table~\ref{tab:boxa}, with a very conservative error given
by the DIS contribution from the region $Q^2 < 1$~GeV$^2$.
This is the value appearing in the column labeled `free nucleons' of
Table~\ref{tab:summary}, together with the neutron contribution using
Eq.~(\ref{eq:pton}).

Because the nucleons in a heavy nucleus are bound, properties
such as their weak charge can differ from those of free nucleons.
In particular, for the elastic $\gz$ box contribution, transitions
to occupied states are forbidden by the Pauli exclusion principle.
To estimate this effect of the nuclear medium on the weak charges,
we consider the expression
\begin{equation}
\!\!\Box_\gz^{N{\rm (el)}} = {2\alpha\over\pi} v_e \mu^N\! g_A^N\!
    \int_{0}^\infty dQ^2\,F_V(Q^2) F_A(Q^2) \, f(Q^2)\, ,
\label{elastic}
\end{equation}
where
\begin{equation}
    f(Q^2) = {1 + 2 \gamma_1 \over Q^2 (1+\gamma_1)^2}, \, \ \ \ \
       \gamma_1 \equiv \gamma(x=1).
\end{equation}
Here $v_e$ is taken at an appropriate low-momentum scale, and the
$Q^2$ dependence of the $Z$ propagator has been dropped.

Pauli blocking is important because the integrand in Eq.~(\ref{elastic})
is heavily weighted towards low $Q^2$,
\begin{equation}
f(Q^2) \xrightarrow{Q^2\rightarrow 0}
{1\over M Q} - {3\over 4 M^2} + \ldots
\label{fdef}
\end{equation}
Since the form factors introduce corrections of order $Q^2/\Lambda^2$,
the dominant low-$Q^2$ contribution is largely independent of nucleon
structure.

To allow for Pauli blocking of the intermediate nucleon in a heavy
nucleus we use the Fermi gas model, where nucleon states are occupied
below the Fermi momentum $k_F$ (typically $\sim 0.25$~GeV).
This estimate should suffice in a heavy nucleus like $\cs$ with a
low surface to volume ratio.  For momentum transfer $\bm{q}$ to a bound
nucleon of momentum $\bm{p}$, we must exclude from the integral of
Eq.~(\ref{elastic}) all intermediate nucleon states of momentum
$|\bm{p}+\bm{q}| < k_F$.
Introducing the occupation number $n_p = \Theta(k_F-p)$, we therefore
have a factor
\begin{equation}
C(q) = {\int d^3 p\, n_p\, n_{|\bm{p}+\bm{q}|}\over \int d^3 p\,n_p}
\end{equation}
to be folded into the integrand of Eq.~(\ref{elastic}),
with $\int d^3 p\,n_p=(4\pi/3) k_F^3$.  This represents the fractional
volume of occupied states that cannot reach the Fermi surface for a
given value of $\bm{q}$.  Since $k_F^2\ll M^2$, the affected states
have nonrelativistic energies, and so $Q^2\approx \bm{q}^2$.
{}From simple geometry, we find
\begin{equation}
C(q) =
\begin{cases}
1 - {1\over 2}\left(3 {q\over 2 k_F}
  - \left({q\over 2 k_F}\right)^3\right), &\!\! 0<q<2 k_F,	\\
0, &\!\! q \ge 2 k_F.
\end{cases}
\end{equation}
The expression $1-C(q)$ is also the Coulomb sum rule for longitudinal
quasielastic scattering in a nonrelativistic Fermi gas.

The Pauli blocking effect can be introduced as a correction
$1-\Delta(k_F)$ to the value of $\Box_\gz^{N \rm(el)}$ in
Eq.~(\ref{elastic}), with
\begin{equation}
\Delta(k_F)
= {\int_{0}^{2 k_F} dQ^2\, F_V(Q^2) F_A(Q^2) \, f(Q^2)\, C(Q) \over
   \int_{0}^\infty  dQ^2\, F_V(Q^2) F_A(Q^2) \, f(Q^2)}
\end{equation}
representing the fractional contribution from the excluded states.
This correction depends on $k_F$, and has only a weak dependence
on nucleon structure through the form factor parameters $\Lambda$,
with other parameters dropping out in the ratio.  In the numerator,
the leading terms are of order ${\cal O}(k_F)$ and ${\cal O}(k_F^2)$,
with the form factor corrections only appearing at order
${\cal O}(k_F^3)$ and higher.

The values of $k_F$ for protons and neutrons are taken from
Hartree-Fock calculations for $^{133}{\rm Cs}$,
which reproduce the experimental charge density \cite{Der00}.
Fitting the Hartree-Fock proton and neutron distributions to a standard
Woods-Saxon form $\rho(r)=\rho_0 / (1+\exp[(r-c)/a])$, normalized to
$Z$ and $N$, respectively, leads to the parameters given in Table~\ref{densityparam}.
From these one can compute the Fermi momentum in the local density
approximation, $\rho = k_F^3/(3\pi^2)$. The central and average values
of $k_F$ are given in Table~\ref{densityparam}.
The latter, which are
used in our calculations, are consistent with those obtained in fits to
experimental quasielastic electron scattering data using a simple Fermi
gas model \cite{Moniz71}.

\begin{table}[t]
\vspace{-6pt}
\begin{center} 
\caption{
  The parameters of the proton and neutron Woods-Saxon densities
  $\rho(r)$ for $^{133}{\rm Cs}$, together with the mean square radii.
  The Fermi momenta $k_F(0)$ are determined from the central nucleon
  densities, with $\langle k_F\rangle$ the corresponding mean.
\label{densityparam}
}
\begin{ruledtabular}
\begin{tabular*}{0.8\columnwidth}{@{\extracolsep{\fill}}lccccc}
& $c$~(fm) & $a$~(fm) & $\langle r^2\rangle$~(fm$^2$) & $k_F(0)$~(GeV) & $\langle k_F\rangle$~(GeV) \\ \hline
$p$ &5.8895 & 0.4010  & 23.03  & 0.241  & 0.218  \\
$n$ &5.9482  & 0.4946  & 24.61  & 0.266  & 0.236  \\
\end{tabular*}
\end{ruledtabular}
\end{center}
\vspace{-12pt}
\end{table}

We find the Pauli blocking correction factor $1-\Delta(k_F)$ is
approximately linear in $k_F$ over the range $0.2-0.3$~GeV, and
well-approximated by the expression
\begin{equation}
1-\Delta(k_F) \approx 0.83 - 1.04\, k_F,
\end{equation}
with $k_F$ in GeV.  Specifically, using the volume-averaged values of
$\langle k_F\rangle$ in Table~\ref{densityparam}, we find a correction
factor of 0.61 to $\Box_\gz^{p(\rm{el})}$ and 0.59 to
$\Box_\gz^{n(\rm{el})}$.  For reasonable values of $k_F$ up to 0.27~GeV
in a very heavy nucleus, the Pauli blocking factors will therefore fall
into the narrow range $0.55-0.60$, suggesting a relatively insignificant
variation in the total value of $\Box_\gz^N$ for nuclei beyond $\cs$.

The effect on $Q_W({\rm Cs})$ is shown in final column of
Table~\ref{tab:summary}.  There is a small shift of $-0.03$ compared
to the free nucleon values, giving $-73.26(4)$.  The individual terms
contributing to the uncertainty are: 
	$\kappa(0) \swms$ ($\pm 0.033$),
	$WW$ boxes ($\pm 0.006$),
	$\rho$ ($\pm 0.016$), and
	$\gz$ ($\pm 0.021$).
Also included in Table~\ref{tab:summary} is our theoretical value for
the weak charge of $\ra$.

In summary, we have computed the effect of $\gz$ exchange corrections
on the weak charge of heavy nuclei such as $\cs$ and $\ra$, using a
recently developed formalism based on dispersion relations and an
expansion of the $\gz$ interference structure function moments.
The results improve earlier estimates based on a quark model
description of the $\gz$ box contributions, allowing a significant
reduction in the theoretical uncertainty.
Compared with the pioneering early estimates of Marciano and Sirlin
\cite{MS}, the new corrections enhance the magnitude of the weak
charge by $\approx 0.16\%$, which is approximately 4 times larger
than the current uncertainty on $\sintwms$, and will therefore
affect future high-precision determinations of the weak angle.

\vspace*{0.5cm}
\begin{acknowledgments}
We thank Jens Erler, Nathan Hall and Marianna Safronova for helpful
discussions.  This work was supported by NSERC (Canada), the DOE contract
No. DE-AC05-06OR23177, under which Jefferson Science Associates, LLC
operates Jefferson Lab, and the Australian Research Council through an
Australian Laureate Fellowship.
\end{acknowledgments}

\vspace*{-0.5cm}

\end{document}